\documentclass[12pt,a4paper]{article}
\begin{document}
\large

\newpage
\begin{center}
{\bf THE UNITED FAMILIES OF MASSIVE NEUTRINOS
OF A DIFFERENT NATURE}
\end{center}
\vspace{1cm}
\begin{center}
{\bf Rasulkhozha S. Sharafiddinov}
\end{center}
\vspace{1cm}
\begin{center}
{\bf Institute of Nuclear Physics, Uzbekistan Academy of Sciences,
Tashkent, 702132 Ulugbek, Uzbekistan}
\end{center}
\vspace{1cm}

At the availability of a nonzero mass, the same neutrino regardless of
whether it refers to Dirac or Majorana fermions, must possess simultaneously
each of the anapole and electric dipole moments. Their interaction with the
field of emission can also lead to the elastic scattering of the longitudinal
polarized neutrinos on a spinless nucleus. Using the cross section of a
process, the united equation has been obtained between the anapole and
electric dipole form factors of Dirac and Majorana neutrinos. It corresponds
in nature to the coexistence of neutrinos of both types. As a consequence,
each Dirac neutrino testifies in favor of the existence of a kind of Majorana
neutrino. They constitute herewith the united families of massive neutrinos
of a different nature. Therefore, any of the earlier measured properties of
neutrinos may serve as a certain indication of the existence simultaneously
of both Dirac and Majorana neutrinos. All findings are also confirmed by the
comparatively new laboratory restrictions on the self-masses of these fermions.
Thereby they state that electromagnetic gauge invariance must have a new
structure, which depends on nature of the inertial mass and says that
P-symmetry of a particle is basically violated
at the expense of its rest mass.

\newpage
The establishment of the nature, similarity and difference of massive
neutrinos of Dirac $(\nu_{D}\neq \overline{\nu}_{D})$
and Majorana $(\nu_{M}=\overline{\nu}_{M})$ types could essentially clear
up the innate properties of elementary particles.
One of the modes of deciding this problem may be a
study of the behavior of these fermions in the nuclear
charge field \cite{1,2,3}.

If neutrinos have a Dirac mass, the possibility of the existence of small
charge \cite{4,5}, magnetic \cite{6}, anapole and electric dipole \cite{3,7}
moments for such particles is not excluded.

In conformity with these ideas of the standard electroweak model, massive
Dirac $(\nu=\nu_{D})$ neutrino interaction with the field of emission may be
described by the electromagnetic current $j_{\mu}$ consisting of the vector
$j_{\mu}^{V}$ and axial-vector $j_{\mu}^{A}$ components:
\begin{equation}
j_{\mu}=j_{\mu}^{V}+j_{\mu}^{A},
\label{1}
\end{equation}
\begin{equation}
j_{\mu}^{V}=\overline{u}(p',s',)[\gamma_{\mu}f_{1\nu}(q^{2})-
i\sigma_{\mu\lambda}q_{\lambda}f_{2\nu}(q^{2})]u(p,s),
\label{2}
\end{equation}
\begin{equation}
j_{\mu}^{A}=\overline{u}(p',s')\gamma_{5}[\gamma_{\mu}
g_{1\nu}(q^{2})-i\sigma_{\mu\lambda}q_{\lambda}g_{2\nu}(q^{2})]u(p,s).
\label{3}
\end{equation}
Here $\sigma_{\mu\lambda}=[\gamma_{\mu},\gamma_{\lambda}]/2,$ $q=p-p',$
$p (p')$ and $s(s')$ denote the 4-momentum and helicity of the incoming
(outgoing) neutrino.

The functions $f_{i\nu_{D}}(q^{2})$ and $g_{i\nu_{D}}(q^{2})$ are the
vector and axial-vector form factors of leptonic current. Of them
$f_{1\nu_{D}}(q^{2}),$ $f_{2\nu_{D}}(q^{2})$ and $g_{2\nu_{D}}(q^{2})$
at zero four-dimensional momentum transfer $(q^{2}=0)$ define the static
values of the neutrino charge and magnetic and electric dipole moments,
for which there are the laboratory and astrophysical estimates \cite{8}.

Insofar as $g_{1\nu_{D}}(0)$ is concerned, it gives \cite{9} the
dimensionless anapole, but its size has not yet been obtained \cite{10}
in the measurements.

As follows from \cite{11}, if neutrinos are truly neutral $(\nu=\nu_{M}),$
the functions $f_{i\nu_{M}}(q^{2})$ and $g_{2\nu_{M}}(q^{2})$ must be equal
to zero. However, according to \cite{12}, for the Majorana neutrino \cite{13}
$f_{i\nu_{M}}(q^{2})$ and $g_{1\nu_{M}}(q^{2})$ are absent, and form factor
$g_{2\nu_{M}}(q^{2})$ has a nonzero value.

Usually it is assumed that $f_{i\nu}(q^{2})$ are C- and P-even.
The function $g_{1\nu}(q^{2})$ must be CP-invariant \cite{9} but
P-noninvariant. Unlike this, $g_{2\nu}(q^{2})$ is C-symmetrical
but CP-odd \cite{14}. Therefore, the axial-vector dipole
moments $g_{i\nu}(q^{2})$ may exist only in the case where
P-invariance is absent.

Returning to (\ref{1}), we remark that the electromagnetic current
$j_{\mu}^{A}$ in many works was presented as
\begin{equation}
j_{\mu}^{A}=\overline{u}(p',s')\gamma_{5}[\gamma_{\mu}q^{2}
G_{1\nu}(q^{2})-i\sigma_{\mu\lambda}q_{\lambda}G_{2\nu}(q^{2})]u(p,s),
\label{4}
\end{equation}
where $G_{1\nu}(0)$ similarly to the form factor $G_{2\nu}(0)$ coincides
with the dimensional size of the corresponding axial-vector moment. Thus,
the definition (\ref{3}) is actually based on the replacements
\begin{equation}
g_{1\nu}(q^{2})=q^{2}G_{1\nu}(q^{2}), \, \, \, \,
g_{2\nu}(q^{2})=G_{2\nu}(q^{2}).
\label{5}
\end{equation}
In these circumstances analysis of the axial-vector processes of
elastic scattering of electrons and neutrinos on nuclei assumed \cite{3}
that both $G_{1\nu}(0)$ and $G_{2\nu}(0)$ may be different from zero even
if neutrinos are electrically neutral. Their structure at $e=|e|$
has the form
\begin{equation}
G_{1\nu_{D}}(0)=\frac{3eG_{F}}{8\pi^{2}\sqrt{2}}, \, \ \, \,
G_{2\nu_{D}}(0)=\frac{3eG_{F}m_{\nu_{D}}}{4\pi^{2}\sqrt{2}},
\label{6}
\end{equation}
\begin{equation}
G_{1\nu_{M}}(0)=\frac{3eG_{F}}{4\pi^{2}\sqrt{2}}, \, \, \, \,
G_{2\nu_{M}}(0)=\frac{3eG_{F}m_{\nu_{M}}}{2\pi^{2}\sqrt{2}}.
\label{7}
\end{equation}
Based on these solutions, one can think that, unlike
$G_{2\nu}(0),$ the value of $G_{1\nu}(0)$ is not connected with the
neutrino mass \cite{3}. Such a difference in the nature
of form factors $G_{1\nu}(0)$ and $G_{2\nu}(0)$ would seem to say
that any CP-symmetrical or CP-antisymmetrical particle
cannot simultaneously have both CP-even anapole
and CP-odd electric dipole moments. This circumstance
becomes more interesting if we take into account that
the same neutrino may not be simultaneously both a
Dirac and a Majorana fermion.

The purpose of this work is to elucidate whether there exists any
connection between the properties of massive neutrinos of a different
nature, and if so what the observed dependence says about the completeness
of their axial-vector picture. We investigate this question by studying the
interaction with a nucleus Coulomb field of the anapole and electric dipole
moments of neutrinos of Dirac $(\nu_{D}=\nu_{e})$ and Majorana
$(\nu_{M}=\nu_{1})$ types on account of longitudinal
polarization of fermions.

The amplitude of such a process may to the lower order in $\alpha$ be
presented as
$$M_{fi}^{em}=\frac{4\pi\alpha}{q^{2}}\overline{u}(p',s')\gamma_{5}
[\gamma_{\mu}g_{1\nu}(q^{2})-i\sigma_{\mu\lambda}q_{\lambda}
g_{2\nu}(q^{2})]\times$$
\begin{equation}
\times u(p,s)<f|J_{\mu}^{\gamma}(q)|i>.
\label{8}
\end{equation}
Here $\nu=\nu_{DL,R}=\nu_{eL,R}$ or $\nu=\nu_{ML,R}=\nu_{1L,R},$
$J_{\mu}^{\gamma}$ is the nuclear target electromagnetic current \cite{15}.

The cross section of elastic scattering of longitudinal polarized massive
neutrinos of a different nature on a spinless nucleus on the basis
of (\ref{8}) can be written in general form as
$$\frac{d\sigma_{em}^{A_{\nu}}(\theta_{\nu},s,s')}{d\Omega}=
\frac{1}{2}\sigma^{\nu}_{o}\{(1+ss')g_{1\nu}^{2}+$$
\begin{equation}
+4m_{\nu}^{2}\eta_{\nu}^{-2}(1-ss')g_{2\nu}^{2}
tg^{2}\frac{\theta_{\nu}}{2}\}F^{2}_{E}(q^{2}),
\label{9}
\end{equation}
where
$$\sigma_{o}^{\nu}=
\frac{\alpha^{2}cos^{2}(\theta_{\nu}/2)}{4E^{2}_{\nu}
(1-\eta^{2}_{\nu})sin^{4}(\theta_{\nu}/2)}, \, \, \, \,
\eta_{\nu}=\frac{m_{\nu}}{E_{\nu}},$$
$$E_{\nu}=\sqrt{p^{2}+m_{\nu}^{2}},\, \, \, \, F_{E}(q^{2})=
ZF_{c}(q^{2}),$$
$$q^{2}=-4E_{\nu}^{2}(1-\eta_{\nu}^{2})sin^{2}\frac{\theta_{\nu}}{2}.$$
Here $\theta_{\nu}$ is the scattering angle, $F_{c}(q^{2})$ is the nucleus
charge $(F_{c}(0)=1)$ form factor, $E_{\nu}$ and $m_{\nu}$ are the neutrino
energy and mass. The index $A_{\nu}$ implies the absence of the interaction
vector part.

The availability of terms $(1+ss')$ and $(1-ss')$ indicates that
an incoming neutrino flux on its passage through the nucleus suffers
elastic scattering either with $(s'=-s)$ or without $(s'=s)$ helicity
flip. Therefore one can reduce the cross section (\ref{9}) to
\begin{equation}
d\sigma_{em}^{A_{\nu}}(\theta_{\nu},s)=
d\sigma_{em}^{A_{\nu}}(\theta_{\nu},g_{1\nu},s)+
d\sigma_{em}^{A_{\nu}}(\theta_{\nu},g_{2\nu},s),
\label{10}
\end{equation}
$$\frac{d\sigma_{em}^{A_{\nu}}(\theta_{\nu},g_{1\nu},s)}{d\Omega}=
\frac{d\sigma_{em}^{A_{\nu}}(\theta_{\nu},g_{1\nu},s'=s)}{d\Omega}=$$
\begin{equation}
=\sigma^{\nu}_{o}g_{1\nu}^{2}F_{E}^{2}(q^{2}),
\label{11}
\end{equation}
$$\frac{d\sigma_{em}^{A_{\nu}}(\theta_{\nu},g_{2\nu},s)}{d\Omega}=
\frac{d\sigma_{em}^{A_{\nu}}(\theta_{\nu},g_{2\nu},s'=-s)}{d\Omega}=$$
\begin{equation}
=4m_{\nu}^{2}\eta_{\nu}^{-2}\sigma^{\nu}_{o}g_{2\nu}^{2}
F_{E}^{2}(q^{2})tg^{2}\frac{\theta_{\nu}}{2}.
\label{12}
\end{equation}
After averaging over $s$ and summing over $s',$ we can present
the cross section (\ref{9}) in the form
\begin{equation}
d\sigma_{em}^{A_{\nu}}(\theta_{\nu})=
d\sigma_{em}^{A_{\nu}}(\theta_{\nu},g_{1\nu})+
d\sigma_{em}^{A_{\nu}}(\theta_{\nu},g_{2\nu}),
\label{13}
\end{equation}
\begin{equation}
\frac{d\sigma_{em}^{A_{\nu}}(\theta_{\nu},g_{1\nu})}{d\Omega}=
\sigma^{\nu}_{o}g_{1\nu}^{2}F_{E}^{2}(q^{2}),
\label{14}
\end{equation}
\begin{equation}
\frac{d\sigma_{em}^{A_{\nu}}(\theta_{\nu},g_{2\nu})}{d\Omega}=
4m_{\nu}^{2}\eta_{\nu}^{-2}\sigma^{\nu}_{o}g_{2\nu}^{2}
F_{E}^{2}(q^{2})tg^{2}\frac{\theta_{\nu}}{2}.
\label{15}
\end{equation}
So it is seen that (\ref{9}) constitutes the two classes of the axial-vector
cross sections, which may symbolically be written as
\begin{equation}
d\sigma_{em}^{A_{\nu}}(\theta_{\nu},s)=
\{d\sigma_{em}^{A_{\nu}}(\theta_{\nu},g_{1\nu},s), \, \, \, \,
d\sigma_{em}^{A_{\nu}}(\theta_{\nu},g_{2\nu},s)\},
\label{16}
\end{equation}
\begin{equation}
d\sigma_{em}^{A_{\nu}}(\theta_{\nu})=
\{d\sigma_{em}^{A_{\nu}}(\theta_{\nu},g_{1\nu}), \, \, \, \,
d\sigma_{em}^{A_{\nu}}(\theta_{\nu},g_{2\nu})\}.
\label{17}
\end{equation}
At the same time it is clear that the cross sections
$d\sigma_{em}^{A_{\nu}}(\theta_{\nu},s)$ and
$d\sigma_{em}^{A_{\nu}}(\theta_{\nu})$ describing the starting processes
with longitudinal polarized and unpolarized neutrinos, respectively,
coincide:
\begin{equation}
\frac{d\sigma_{em}^{A_{\nu}}(\theta_{\nu},s)}
{d\sigma_{em}^{A_{\nu}}(\theta_{\nu})}=1.
\label{18}
\end{equation}
This ratio together with (\ref{10}) and (\ref{13}) implies that regardless
of the sizes of cross sections, the interratio of any pair of elements from
the sets (\ref{16}) and (\ref{17}) for $\nu_{D}$ and $\nu_{M}$ has the same
value and thereby allows us to establish the six different equations.

But here we will use only one of them, namely
\begin{equation}
\frac{d\sigma_{em}^{A_{\nu_{D}}}(\theta_{\nu_{D}},g_{2\nu_{D}})}
{d\sigma_{em}^{A_{\nu_{D}}}(\theta_{\nu_{D}},g_{1\nu_{D}})}=
\frac{d\sigma_{em}^{A_{\nu_{M}}}(\theta_{\nu_{M}},g_{2\nu_{M}})}
{d\sigma_{em}^{A_{\nu_{M}}}(\theta_{\nu_{M}},g_{1\nu_{M}})},
\label{19}
\end{equation}
which does not depend on the spin phenomena. This equality is a consequence
of the connection that the functions $g_{1\nu}(q^{2})$ and $g_{2\nu}(q^{2})$
describe the most diverse forms of the same regularity of the axial-vector
nature of the neutrino. In other words, it is valid only for an unpolarized
particle possessing simultaneously each of the anapole and electric dipole
moments.

Inserting the corresponding values of cross sections from (\ref{14}) and
(\ref{15}) in (\ref{19}), we are led to an implication that
\begin{equation}
m_{\nu_{D}}^{2}\frac{g_{2\nu_{D}}^{2}(q^{2})}{g_{1\nu_{D}}^{2}(q^{2})}
\frac{tg^{2}\frac{\theta_{\nu_{D}}}{2}}{\eta_{\nu_{D}}^{2}}=
m_{\nu_{M}}^{2}\frac{g_{2\nu_{M}}^{2}(q^{2})}{g_{1\nu_{M}}^{2}(q^{2})}
\frac{tg^{2}\frac{\theta_{\nu_{M}}}{2}}{\eta_{\nu_{M}}^{2}}.
\label{20}
\end{equation}
If we take into account that for the case of $E_{\nu}\gg m_{\nu}$ when
$\eta_{\nu}\rightarrow 0,$ the limit $q^{2}\rightarrow 0$ may take
place only at $\theta_{\nu}\rightarrow 0,$ jointly with the size
$$lim_{\eta_{\nu}\rightarrow 0,\theta_{\nu}\rightarrow 0}
\frac{tg^{2}\frac{\theta_{\nu}}{2}}{\eta_{\nu}^{2}}=\frac{1}{4},$$
the equality (\ref{20}) states that
\begin{equation}
m_{\nu_{D}}\frac{g_{2\nu_{D}}(0)}{g_{1\nu_{D}}(0)}=
\pm m_{\nu_{M}}\frac{g_{2\nu_{M}}(0)}{g_{1\nu_{M}}(0)}.
\label{21}
\end{equation}
We see that between the axial-vector properties of massive Dirac and
Majorana neutrinos there exists a sharp dependence. This connection
corresponds in nature to the coexistence of neutrinos of both types.
Therefore, from its point of view, it should be expected that each Dirac
$(\nu_{D}=\nu_{e}, \nu_{\mu}, \nu_{\tau}, ...)$ neutrino testifies in favor
of the existence of a kind of the Majorana
$(\nu_{M}=\nu_{1}, \nu_{2}, \nu_{3}, ...)$ neutrino. They constitute
herewith the united families of massive neutrinos of a different nature:
\begin{equation}
\pmatrix{\nu_{e}\cr \nu_{1}}, \, \, \, \,
\pmatrix{\nu_{\mu}\cr \nu_{2}}, \, \, \, \,
\pmatrix{\nu_{\tau}\cr \nu_{3}}, ....
\label{22}
\end{equation}

As well as in the systems of leptons and hadrons, each purely neutrino
family here must distinguish itself from others by the individual properties
of a pair of structural particles.

In particular, the presence of the term $(1-ss')$ in (\ref{9}) should be
recalled, describing the fact that owing to an intimate connection between
the mass of the neutrino and its physical nature, any of the interconversions
$\nu_{L}\leftrightarrow \nu_{R}$ and
$\overline{\nu}_{R}\leftrightarrow \overline{\nu}_{L}$ in the field of
a nucleus becomes fully possible. However, in the framework of the standard
electroweak theory, their existence is incompatible with the chiral symmetries
of elementary particles.

Analysis of such transitions assumed that the chirality of the neutrino
is basically violated at the expense of the flip of its spin \cite{16} arising
from the availability of the neutrino nonzero rest mass \cite{4}. Therefore,
to establish the legality of the discussed procedure and to use it in the
description of the neutrino interaction with a field of emission, one must
elucidate the nature of the corresponding mass responsible for the distinction
of each of family (\ref{22}) among other purely neutrino, leptonic
and hadronic doublets.

But at a given stage we can only add that the suggested family structure
(\ref{22}) that unites the massive Dirac and Majorana neutrinos takes place
owing to the unified nature of these fermions, which means that the absence
of one of them, as stated in (\ref{21}), would imply that neither exists at
all. From this point of view, the Cowan and Reines \cite{17} experiments may
serve as the first confirmation of the existence simultaneously of both the
Dirac and Majorana neutrinos but not of one of them.

Furthermore, if we suppose that $m_{\nu_{D}}=0,$ then (\ref{21})
would lead us to an implication about the equality of
a truly neutral neutrino Majorana rest mass to zero.
There exist, however, comparatively new laboratory
restrictions on the self-masses \cite{18,19} of neutrinos
of both types.

Insofar as their interratio is concerned, it is in favor of the united
connection (\ref{21}), which states that
\begin{equation}
\frac{m_{\nu_{D}}}{m_{\nu_{M}}}=
\frac{g_{1\nu_{D}}(0)}{g_{1\nu_{M}}(0)}
\frac{g_{2\nu_{M}}(0)}{g_{2\nu_{D}}(0)}.
\label{23}
\end{equation}

Another consequence of the structural dependence (\ref{21}) is that
\begin{equation}
m_{\nu_{D}}g_{1\nu_{M}}(0)g_{2\nu_{D}}(0)-
m_{\nu_{M}}g_{1\nu_{D}}(0)g_{2\nu_{M}}(0)=0.
\label{24}
\end{equation}

In conformity with the ideas of the nature of the
micro-world symmetry, any massive C-invariant or C-
noninvariat neutrino may not be simultaneously both
a Dirac and a Majorana particle. According to this
point of view, each interference term in (\ref{24})
implies the existence of a kind of unified system of
the two types of neutrinos.

To elucidate the compound structure of such pairs one must apply once more
to the cross section (\ref{10}), from which it follows \cite{20} that the
anapole $g_{1\nu}$ is responsible for the scattering without flip of spin,
and the electric dipole moment $g_{2\nu}$ answers to the interconversions
of neutrinos of the different helicities. Of course, these transitions can
also confirm the fact that the same neutrino may not have simultaneously both
left-handed and right-handed fermions. It is fully possible therefore that
any interference between the two interactions of a different CP-invariance
corresponds in nature to an individual system of the left-
and right-handed or the two left (right)-handed particles
of the same families of doublets and singlets.

It is already clear from the above that each of (\ref{21})
and (\ref{24}) describes the naturally united system of the
two individual systems of the left- and right-handed
or the two left (right)-handed neutrinos of the most
diverse currents.

Such a connected system of the four types of neutrinos
appears as a consequence of symmetry laws of
elementary particles, even with elastic scattering on a
spinless nucleus \cite{21}, if an incoming neutrino flux in it
includes both Dirac and Majorana fermions. Therefore
a subtle measurement \cite{18,19} of the neutrino self-mass
one must use as the one of the available experimental
data sets confirming the existence of right-handed
neutrinos of a different nature and that the
number of components in all types of massive particles
must coincide. This makes it possible to understand the nature of
the above-discussed purely neutrino family of fermions at
the fundamental dynamical level. In other words, massive
Dirac and Majorana neutrinos
constitute the naturally united families not only of the left-handed
$SU(2)_{L}$-doublets but also of the right-handed $SU(2)_{R}$-singlets.
Formulating more concretely, one can define their
structure in general form:
\begin{equation}
\pmatrix{\nu_{e}\cr \nu_{1}}_{L},
(\nu_{e}, \, \, \, \, \nu_{1})_{R}, \, \, \, \,
\pmatrix{\nu_{\mu}\cr \nu_{2}}_{L},
(\nu_{\mu}, \, \, \, \, \nu_{2})_{R}, \, \, \, \,
\pmatrix{\nu_{\tau}\cr \nu_{3}}_{L},
(\nu_{\tau}, \, \, \, \, \nu_{3})_{R}, ....
\label{25}
\end{equation}

For completeness we remark that insertion of (\ref{5}) in (\ref{20})
suggests
\begin{equation}
\frac{G_{2\nu_{D}}^{2}(q^{2})}{4m_{\nu_{D}}^{2}G_{1\nu_{D}}^{2}(q^{2})}
\frac{\eta_{\nu_{D}}^{2}tg^{2}\frac{\theta_{\nu_{D}}}{2}}
{(1-\eta_{\nu_{D}}^{2})^{2}sin^{4}\frac{\theta_{\nu_{D}}}{2}}=
\frac{G_{2\nu_{M}}^{2}(q^{2})}{4m_{\nu_{M}}^{2}G_{1\nu_{M}}^{2}(q^{2})}
\frac{\eta_{\nu_{M}}^{2}tg^{2}\frac{\theta_{\nu_{M}}}{2}}
{(1-\eta_{\nu_{M}}^{2})^{2}sin^{4}\frac{\theta_{\nu_{M}}}{2}}.
\label{26}
\end{equation}
By following the same arguments that led to solution (\ref{21}), but having
in view the equality
$$lim_{\eta_{\nu}\rightarrow 0,\theta_{\nu}\rightarrow 0}
\frac{\eta_{\nu}^{2}tg^{2}\frac{\theta_{\nu}}{2}}
{(1-\eta_{\nu}^{2})^{2}sin^{4}\frac{\theta_{\nu}}{2}}=1,$$
one can found from (\ref{26}) that
\begin{equation}
\frac{G_{2\nu_{D}}(0)}{2m_{\nu_{D}}G_{1\nu_{D}}(0)}=
\pm \frac{G_{2\nu_{M}}(0)}{2m_{\nu_{M}}G_{1\nu_{M}}(0)}.
\label{27}
\end{equation}
Its comparison with (\ref{21}) at $g_{2\nu}(0)=G_{2\nu}(0)$ allows
one to conclude that
\begin{equation}
g_{1\nu}(0)=2m_{\nu}^{2}G_{1\nu}(0).
\label{28}
\end{equation}
Uniting (\ref{28}) with the corresponding sizes from (\ref{6}) and (\ref{7}),
we get the following dependences of form factors
\begin{equation}
g_{1\nu_{D}}(0)=
\frac{3eG_{F}m_{\nu_{D}}^{2}}{4\pi^{2}\sqrt{2}}, \, \ \, \,
g_{1\nu_{M}}(0)=\frac{3eG_{F}m_{\nu_{M}}^{2}}{2\pi^{2}\sqrt{2}}.
\label{29}
\end{equation}
If we now take into account that the existence of an intimate connection
between the mass of the neutrino and its axial-vector nature is by no means
excluded naturally, then there arises a question of whether the absence of
any of the form factors $g_{1\nu}(0)$ or $g_{2\nu}(0)$ is not strictly
nonverisimilar even at the violation of CPT-symmetry of a particle itself.

Here it is relevant to note that the term $g_{1\nu}(q^{2})$ in (\ref{3})
is incompatible with electromagnetic gauge invariance, the conservation of which
would imply its equality to zero. However, according to (\ref{29}), this
would take place only in the case of the neutrino mass
being wholly absent. Therefore without contradicting
implications of the united equation (\ref{21}), we conclude
that a connection between $g_{1\nu}(0)$ and $m_{\nu}$ testifies in
favor of a new regularity of the compound structure
of gauge invariance \cite{5}, which depends on a particle
mass and says that on the availability of a nonzero
mass, the same neutrino regardless of whether it
refers to Dirac or Majorana fermions, must possess
simultaneously both CP-even anapole and CP-odd electric
dipole moments, if the P-symmetry of each of
these currents is basically violated at the expense
of the neutrino mass.

Finally, insofar as an experimental possibility of establishing the nature
of the discussed types of neutrino interactions is concerned, we will start
from the fact \cite{22} that the axial-vector electromagnetic scattering
angle $\theta_{\nu}$ may be defined as a function of the neutrino anapole
$g_{1\nu}(q^{2})$ and electric dipole $g_{2\nu}(q^{2})$ form factors:
\begin{equation}
\theta_{\nu}=\pm 2Arctg\frac{g_{1\nu}(q^{2})}{2E_{\nu}g_{2\nu}(q^{2})}.
\label{30}
\end{equation}
It is clear, however, that high energy fermions $(E_{\nu}\gg m_{\nu})$
for which $\eta_{\nu}\rightarrow 0,$ suffer the scattering almost forward,
namely $\theta_{\nu}\rightarrow 0.$

In these circumstances $q^{2}\rightarrow 0,$ and cross section of the
elastic scattering of the unpolarized neutrinos (\ref{13}) on account
of $g_{1\nu}^{2}(0)/E_{\nu}^{2}\rightarrow 0$ has size
\begin{equation}
\frac{d\sigma_{em}^{A_{\nu}}(\theta_{\nu})}{d\Omega}=
\frac{Z^{2}\alpha^{2}}{sin^{2}(\theta_{\nu}/{2})}g_{2\nu}^{2}(0).
\label{31}
\end{equation}
Thus it follows that measurement of $\theta_{\nu}$ for the two different
values of large energies allows us to estimate the neutrino anapole and
electric dipole moments \cite{22}. Thereby they make it possible to define
the cross section (\ref{31}) as well as the fine-structure constant $\alpha,$
whose size has not yet been found in the neutrino experiments.

For such measurements it is desirable to choose only the one of spinless
nuclei with zero isospin (for example, ($^{4}He,$ $^{12}C,$ $^{40}Ca,$ ...)),
so that the target nucleus isotopic structure can essentially influence the
neutrino electromagnetic properties \cite{23}. Of course, the above-noted
regularities of the axial-vector nature of massive neutrinos take place
regardless of the choice of conditions for their observations.

\end{document}